# Ternary Tetradymite Compounds as Topological Insulators

Lin-Lin Wang[1*] and Duane D. Johnson[1,2§]

[1]Division of Materials Science and Engineering, Ames Laboratory, Iowa State University, Ames, Iowa 50011; [2]Department of Materials Science and Engineering, Iowa State University, Ames, IA 50011

## Abstract

Ternary tetradymites $Bi_2Te_2S$, $Bi_2Te_2Se$ and $Bi_2Se_2Te$ are found to be stable, bulk topological insulators via theory, showing band inversion between group V and VI $p_z$ orbitals. We identify $Bi_2Se_2Te$ as a good candidate to study massive Dirac Fermions, with a (111) cleavage-surface-derived Dirac point (DP) isolated in the bulk band gap at the Fermi energy $E_f$ like $Bi_2Se_3$ but with a spin texture alterable by layer chemistry. In contrast, $Bi_2Te_2S$ and $Bi_2Te_2Se$ (111) behave like $Bi_2Te_3$, with a DP below $E_f$ buried in bulk bands. $Bi_2Te_2S$ offers large bulk resistivity needed for devices.

Materials that exhibit topological insulator (TI) behavior reveal a novel quantum state for electrons,[1,2] where surface states of a 3-dimensional (3D) system are topologically protected against disorder by time-reversal symmetry (TRS) and, as a result, electrons experiences no backward scattering by non-magnetic impurities. The unique feature of a 3D TI lies in its band dispersion – surface bands connect valence and conduction bands and cross the Fermi level $E_f$ an odd number of times along two TRS-equivalent k-points, typically found in narrow-gap semiconductors with strong spin-orbit coupling (SOC). Since $Bi_2Se_3$ and $Bi_2Te_3$ were concurrently observed[3-5] and predicted[6] to be 3D TI, an intensive search[7-10] continues for other systems. Notably, $Bi_2Se_3$ and $Bi_2Te_3$ belong to a class of line compounds called *tetradymites*. Ternary tetradymites, such as $Bi_2Te_2S$, $Bi_2Te_2Se$ and $Bi_2Se_2Te$, are also stable[11,12] and potentially offer a "chemistry knob" to control TI behavior. Here, we provide insight for device development – cleavage surface-band dispersion and location of Dirac point (DP) both relative to bulk bands, as well as the warping (out-of-plane spin component, referred hereafter as *spin texture*) of the Dirac cone (DC).

Upon band inversion and crossing of $E_f$ due to SOC, a DP is formed leading to variety of unique physics. Using angle-resolved photoelectron spectroscopy, Chen et al.[13] have shown that massive Dirac Fermion is produced on $Bi_2Se_3$ (111) surface by breaking TRS via magnetic impurities. To do this, the position of DP must be in the gaps of both bulk and surface bands. $Bi_2Se_3$ has a DP isolated in the bulk-band gap region making it a good candidate for such studies, whereas the DP for $Bi_2Te_3$ is below $E_f$ and buried among surface-valence bands (Fig. 4 in Ref. 6). Also, $Bi_2Se_3$ has a larger band gap compared to $Bi_2Te_3$, offering more control at elevated temperature. Another large difference between the binary tetradymites is the shape of DC. In contrast to the perfect DC of $Bi_2Se_3$, $Bi_2Te_3$ exhibits warping and a spin texture,[14,15] suggested to form exotic charge and spin-density waves. Importantly, then, it is more helpful for theory to predict not only TI but also to detail surface-band dispersion.

Although the calculated[7] bulk bands of $Bi_2Te_2Se$ and $Bi_2Te_2S$ indicate TI behavior, with $Bi_2Te_2Se$ confirmed experimentally,[16] the positions of surface-derived DP and DC features (spin texture) have not been analyzed. Also, there has been no study on $Bi_2Se_2Te$. We focus on bulk and surface band TI features of these ternary tetradymites as calculated in density functional theory[17,18] (DFT). We analyze both the bulk and surface (via a slab model) band structures to identify any 3D TIs along with the nature of the DP, especially its location and shape of DC. We find that all of them are 3D TIs, with $Bi_2Te_2S$ and $Bi_2Te_2Se$ behaving like $Bi_2Te_3$, but with higher impurity formation energy than $Bi_2Te_3$ suggesting lower bulk conductivity. We also present band dispersion for two TI $Bi_2Se_2Te$ structures (a variant of a partially-disorder structure, verified as the lowest-energy structure, and a hypothetical structure) to address the effect of site substitution (chemistry) on the shape of DC and spin texture.

One costly method to tell a TI from an ordinary band insulator is to calculate the $Z_2$ topological order parameter.[19] Yet, for structures with inversion symmetry, Fu et al.[20] proposed inspecting the parity product of occupied bands on TRS equivalent $k$-points in the bulk bands. Here we search for a band inversion and zero gap in bulk bands by tuning the SOC ($\lambda$ from 0 to 100%), and then, once verified, we calculate (slab) surface bands and assess DP formation, an approach successfully used for semi-infinite surfaces of the binaries by Zhang et al.[6]

Tetradymite compounds formed between group V and VI elements have a quintuple-layered structure, in which group VI element occupies the outmost and central (third) layer, and group V element occupies the second layer, e.g., for $Bi_2Se_3$ the layers stack as $Se^I$-$Bi$-$Se^{II}$-$Bi$-$Se^I$ along <111> in the primitive rhombohedral cell with the space group of $R\bar{3}m$ (No. 166). The two group VI positions are not equivalent. The stacking is similar to fcc, but the interlayer distances are different; in particular, the distance between neighboring quintuple-layer units is the largest, making it easier to cleave with group VI element exposed as the surface layer. We use DFT with PW91 exchange-correlation functional[21] and plane-wave basis set with projected augmented waves,[22] as implemented in VASP.[23,24] Bulk tetradymite can also be represented as a hexagonal lattice with fifteen atomic layers, which also gives the basis in the slab model for the (111) surface. Both the atomic structure and Brillouin zone for tetradymite bulk and (111) surface have been shown before,[6,25] so we do not repeat them. For bulk we use the primitive rhombohedral cell of five atoms with $7\times7\times7$ $k$-point mesh. We use a three quintuple-layered slab along <111> with no vacuum for bulk band projection and a 14 Å vacuum for surface-band calculations. The kinetic energy cutoff is 280 eV. The $k$-point meshes used are $10\times10\times2$ and $10\times10\times1$, respectively. The convergence with respect to k-point mesh was carefully checked, with total energy converged, e.g., well below 1 $m$eV/atom. We use experimental lattice constants (Table 1) and fix atoms in their bulk positions. To show a TI surface band connects conduction and valence bands, a semi-infinite surface is required[6] and achieved with a thick slab model.[26] At least three quintuple layers along <111> are needed to model the band dispersion in semi-infinite surface with the group VI terminated surfaces well separated by 28 Å. We provide validation of our surface-band structure for binary $Bi_2Se_3$ and $Bi_2Te_3$ in the supporting materials.

In tetradymites, central-layer substitution is commonly observed, e.g., $Bi_2Te_2S$ and $Bi_2Te_2Se$.[11,12] For $Bi_2Se_2Te$, the structure with Te substituting for Se in the central layer of $Bi_2Se_3$ (we call $Bi_2Se_2Te$-I) has not been observed. However, we have calculated that a partially-disordered structure (we call $Bi_2Se_2Te$-II) suggested from X-ray diffraction[12] is a very low-energy structure, in which half of the Se in the outer layers are replaced randomly with Te, which conforms to Hume-Rothery solid-solution rules. To assess behavior, we used an ordered (2×2) supercell approximate for the partially-disordered structure. Because our focus is on electronic structure and TI behavior, we include both $Bi_2Se_2Te$-I and $Bi_2Se_2Te$-II and show the chemistry effects on band dispersion. The details in structural complexity, energetics and phase diagram of $Bi_2Se_3$-$Bi_2Te_3$ will be reported later.[27]

For the prototype ternary tetradymite, $Bi_2Te_2S$, where Te in the central layer of $Bi_2Te_3$ is replaced with S, a slightly smaller lattice constant is found, see Table 1. The bulk band, Fig. 1(c), shows an indirect gap along high symmetry directions ($\Delta_{hs}$) of 0.28 eV. We also calculate the band gap at $\Gamma$ point ($\Delta_\Gamma$) vs. $\lambda$ (with charge density fixed at that of $\lambda$=0). As seen in Fig. 1(d), $\Delta_\Gamma$ becomes 0 around the critical value $\lambda_c$=44%. Compared to the band structure without SOC in Fig. 1(a), the largest change is the lowering of the lowest-conduction band and the rising of the highest-valence band to form jointly a DP at $\Gamma$ point at $E_f$. Upon further increase of $\lambda$, Fig. 1(c) shows that the gap at $\Gamma$ point is reopened and two nearby maximum are formed. Fig. 1(e) shows the projections of the lowest-conduction band on Bi and Te $p_z$ orbital vs. $\lambda$. Below (above) 44% SOC, it is mostly composed of Bi (Te) $p_z$ components. Indeed, SOC causes the band inversion in $Bi_2Te_2S$, similar to the binaries,[6] so $Bi_2Te_2S$ is also a 3D TI. The data in Table 1 may be useful for device development.

To see the surface bands of $Bi_2Te_2S(111)$, we plot in Fig. 1(f) the dispersion of a three quintuple slab on top of the projected bulk bands. Similar to $Bi_2Te_3$ (see Fig. S1(d)), we find the DP for Bi$_2$Te$_2$S(111) is below $E_f$ and buried by other states around the $\Gamma$ point, showing clearly that the two outmost surface layers in the tetradymite mostly determine the position of DP.

Ternary Bi$_2$Te$_2$Se is similar to Bi$_2$Te$_2$S, with the central Te layer in Bi$_2$Te$_3$ replaced by Se. In Fig. 2(a) and (b) we show its bulk and surface band dispersion. By tuning $\lambda$, we find that the bulk band inversion occurs when $\lambda_c$=21% (Table 1). Thus, Bi$_2$Te$_2$Se is also a 3D TI, with a $\Delta_{hs}$ of 0.28 eV – larger than that of Bi$_2$Te$_3$, mostly due to a smaller lattice constant. Looking at slab surface bands, we find similar features to that of Bi$_2$Te$_3$ for the position of DP; i.e., the DP of Bi$_2$Te$_2$Se is buried deeper below $E_f$ (−0.08 eV) than that of Bi$_2$Te$_3$ (−0.02 eV) and Bi$_2$Te$_2$S (−0.05 eV), see Fig. 2.

Fig. 2 (c)-(d) show bulk and surface dispersions for the hypothetical Bi$_2$Se$_2$Te-I, where Te replaces Se in the central layer of Bi$_2$Se$_3$. At $\lambda_c$=71% a band inversion at the $\Gamma$ point for bulk indicates that it is a 3D TI. The surface bands in Fig. 2(d) shows that it behaves much like Bi$_2$Se$_3$ with the DP located in the gap region and not buried by other valence bands. In contrast, if Te substitutes instead in the outer layers, consistent with Hume-Rothery's rules, we find the Bi$_2$Se$_2$Te-II surface band structure in Fig. 2(e), where the highest-valance band around $\Gamma$ point becomes more flat (beyond the effect of zone folding) and the isolation of the DP in the gap region remains but is less prominent than for Bi$_2$Se$_2$Te-I. This direct comparison again shows that the outer two surface layers mostly determined the locations of the valence band and DP.

A significant difference between the surface band structures of Bi$_2$Se$_3$ and Bi$_2$Te$_3$ is the shape of DC and the associated spin texture. Figure 3 shows the shape of the conduction band of each compound around the $\Gamma$ point. Comparing Fig. 3(a) and (d), the DC for Bi$_2$Te$_3$ remains ideal up to 0.2 eV above $E_f$, while for Bi$_2$Se$_3$ it is 0.4 eV, agreeing with previous experiments[5,13] and theory.[14,15] Above these energies, there is significant warping of the DC for Bi$_2$Te$_3$, as evidenced by the change in the shape of contour lines on the bottom – first to a hexagon, then to a snowflake. Such nonconvex shapes produce more pairs of stationary points on the constant-energy contours, allowing scattering processes among different pairs of stationary points. This result is in contrast to no scattering for a convex DC as observed in STM experiment,[28] where a line defect on Bi$_2$Te$_3$(111) suppresses scattering only in the energy range of circular constant-energy contours, not snowflake types.

Figure 3 also shows the spin texture associated with the DC, i.e., the ratio of out-of-plane to total electron-spin moment color-mapped on the cone. For a perfect DC, the electron spin always lies in the surface plane and is perpendicular to the wave vector. In contrast, for a warped DC, Fu[14] suggested that there should be a significant amount of out-of-plane spin moments (up to 60%) to maintain a Berry phase change of $\pi$ in one circuit, as required by topological invariance. Such behavior can lead to interesting features, such as spin-density waves on a 3D TI surface and opening of DP by an in-plane magnetic field. For $Bi_2Te_3$, our results in Fig. 3(a) agree with a previous calculation[15] and show a large spin texture above 0.2 eV, except for $\overline{\Gamma}-\overline{M}$. In contrast, $Bi_2Se_3$ has a much smaller spin texture, Fig. 3(d). For Te-rich ternaries, compared to $Bi_2Te_3$, the upper warping limit of the cone convexity for $Bi_2Te_2S$ and $Bi_2Te_2Se$ are both increased to 0.3 eV, see Fig. 3(b) and (c), and beyond that only a small spin textures appear, with $Bi_2Te_2Se$ being larger than $Bi_2Te_2S$. On the Se-rich side, compared to $Bi_2Se_3$, the upper warping limit of the cone convexity for $Bi_2Se_2Te$-II, see Fig. 3(e), is slightly decreased and beyond that a slightly larger spin texture appears; in contrast, for $Bi_2Se_2Te$-I, see Fig. 3(f), the upper warping limit is reduced to 0.3 eV and a much larger spin texture appears, similar to $Bi_2Te_3$. This shows that, even though the position of DP and dispersion of valence bands are only slightly affected by the substitution of atom in the central layer (Fig. 2), the DC warping and spin texture in conduction band are greatly affected by chemical substitutions. The change in the warping limit reflects the stronger trigonal crystal potential of Te than Se and S.

In $Bi_2Te_2Se$ and $Bi_2Se_2S$, the binding of the site in the central layer to two Bi sites increases hybridization and decreases spin-orbit coupling. While in $Bi_2Se_2Te$-I, by replacing the central Se layer with Te, the spin-orbit coupling in the Bi-Te-Bi trilayer is enhanced significantly and results in a large spin texture. Thus, introducing Te into the central layer of $Bi_2Se_3$ can make a superb TI candidate with a DP standing alone in bulk band gap (just like $Bi_2Se_3$) as well as a large spin texture (just like $Bi_2Te_3$ and larger than $Bi_2Se_3$). But finding feasible ways to introduce Te into the central layer poses a challenge. However, while spin texture in $Bi_2Se_2Te$-II is not as large as that in $Bi_2Se_2Te$-I, it is larger than $Bi_2Se_3$. All other electronic features are similar, including isolation of the DP in the gap, suggesting the lower-energy partially-disordered structure will also be a superb TI candidate.

Lastly, any potential use of these materials as a 3D-TI device requires control over the bulk resistivity. Ideal binary tetradymites $Bi_2Se_3$ and $Bi_2Te_3$ are semiconductors, but defects (such as vacancy and Bi-Te-antisite) cause significant bulk conductivity, which overwhelms the surface-state contribution. For example, Ren et al.[16] have shown experimentally that the ternary $Bi_2Te_2Se$ has a much larger bulk resistivity than $Bi_2Te_3$ because the substitution of the central Te site with Se reduces the formation of a Se vacancy and Bi-Te antisite. The same mechanism should also be operative for $Bi_2Te_2S$ because S is more electronegative and binds Bi stronger than Se, preserving the stoichiometric structure even better than $Bi_2Te_2Se$. For example, the energy cost to create a Bi-Te-antisite pair is increased by 0.06 eV when changing from $Bi_2Te_2Se$ to $Bi_2Te_2S$.[27] A similar situation would hold for quaternary $Bi_2(Te-Se)_2S$.

In conclusion, we find that $Bi_2Te_2S$, $Bi_2Te_2Se$ and $Bi_2Se_2Te$ ternary tetradymites are bulk topological insulators, confirmed computationally by verifying band inversion between group V and VI $p_z$ orbitals. We validated and then used band structures of a large (three quintuple-layered) slab model to study surface-band features, including the warping of DC and its associated spin texture. The location of the DP for $Bi_2Te_2Se$ and $Bi_2Te_2S$ (111) surfaces is like that of $Bi_2Te_3$. We also present dispersion for the energetically preferred *partially-disordered* $Bi_2Se_2Te$ (via an ordered approximate) that showed it had a favorably located surface-derived DP near $E_f$ in the bulk band gap like $Bi_2Se_3$ but with a larger spin texture. We studied a hypothetical $Bi_2Se_2Te$ structure to show that substitution in the central layer affects the shape of DC and the associated spin texture of conduction band, while substitution in outer layers affects the location of the DP and dispersion of valence band. Due to high defect-formation energy, $Bi_2Te_2S$ should have a large bulk resistivity needed to realize a workable device.

Work at Ames Laboratory was supported by the U.S. Department of Energy, Office of Basic Energy Sciences, Division of Materials Science and Engineering. Ames Laboratory is operated for DoE by Iowa State University under Contract No. DE-AC02-07CH11358.

* llw@ameslab.gov, § ddj@ameslab.gov

|  | $a$ (Å) | $c$ (Å) | $x_1$ | $x_2$ | $\lambda_c$ | $\Delta_{hs}$ (eV) | $\Delta_{\Gamma}$ (eV) |
|---|---|---|---|---|---|---|---|
| $Bi_2Se_3$ | 4.138 | 28.64 | 0.399 | 0.206 | 0.46 | 0.32 | 0.47 |
| $Bi_2Te_3$ | 4.383 | 30.487 | 0.400 | 0.212 | 0.48 | 0.14 | 0.52 |
| $Bi_2Se_2Te$ | 4.218 | 29.240 | 0.398 | 0.211 | 0.71 | 0.17 | 0.25 |
| $Bi_2Te_2Se$ | 4.28 | 29.86 | 0.396 | 0.211 | 0.21 | 0.28 | 0.70 |
| $Bi_2Te_2S$ | 4.316 | 30.01 | 0.392 | 0.212 | 0.44 | 0.28 | 0.53 |

Table 1. Experimental (Ref. 11 and 12) structural parameters (lattice constant $a$, $c$ and internal parameters $x_1$ and $x_2$) and DFT-PW91 results for bulk tetradymite compounds, i.e., critical strength of SOC for band inversion at $\Gamma$ point ($\lambda_c$), lowest band gap along high symmetry directions ($\Delta_{hs}$) considered and band gap at $\Gamma$ point ($\Delta_{\Gamma}$).

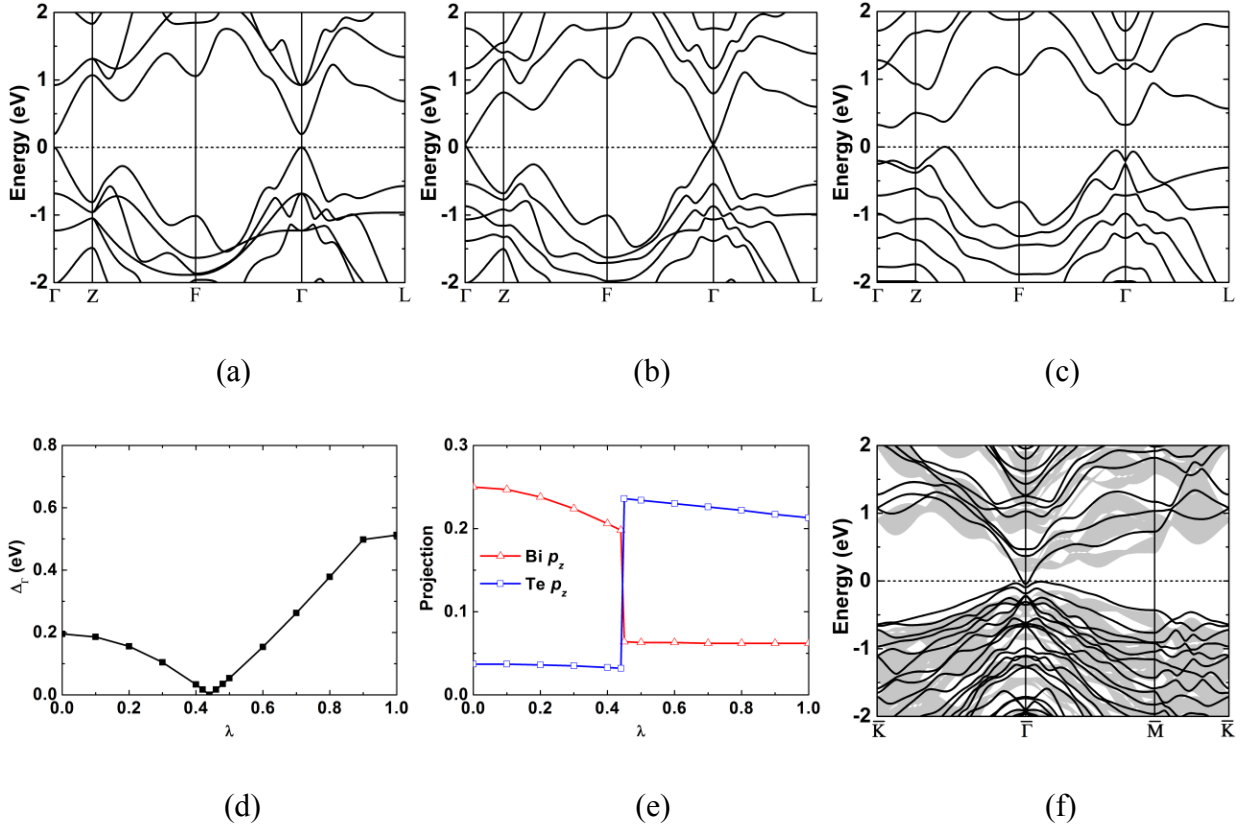

Fig. 1. Band structures of bulk $Bi_2Te_2S$ are shown with SOC strength (λ) of 0, 44 and 100% in (a), (b) and (c), respectively. (d) Bulk band gap at $\Gamma$ point vs. $\lambda$. (e) Projection of the lowest conduction band at $\Gamma$ point on Bi and Te $p_z$ orbital vs. $\lambda$. (f) Band structure of a $Bi_2Te_2S$ slab.

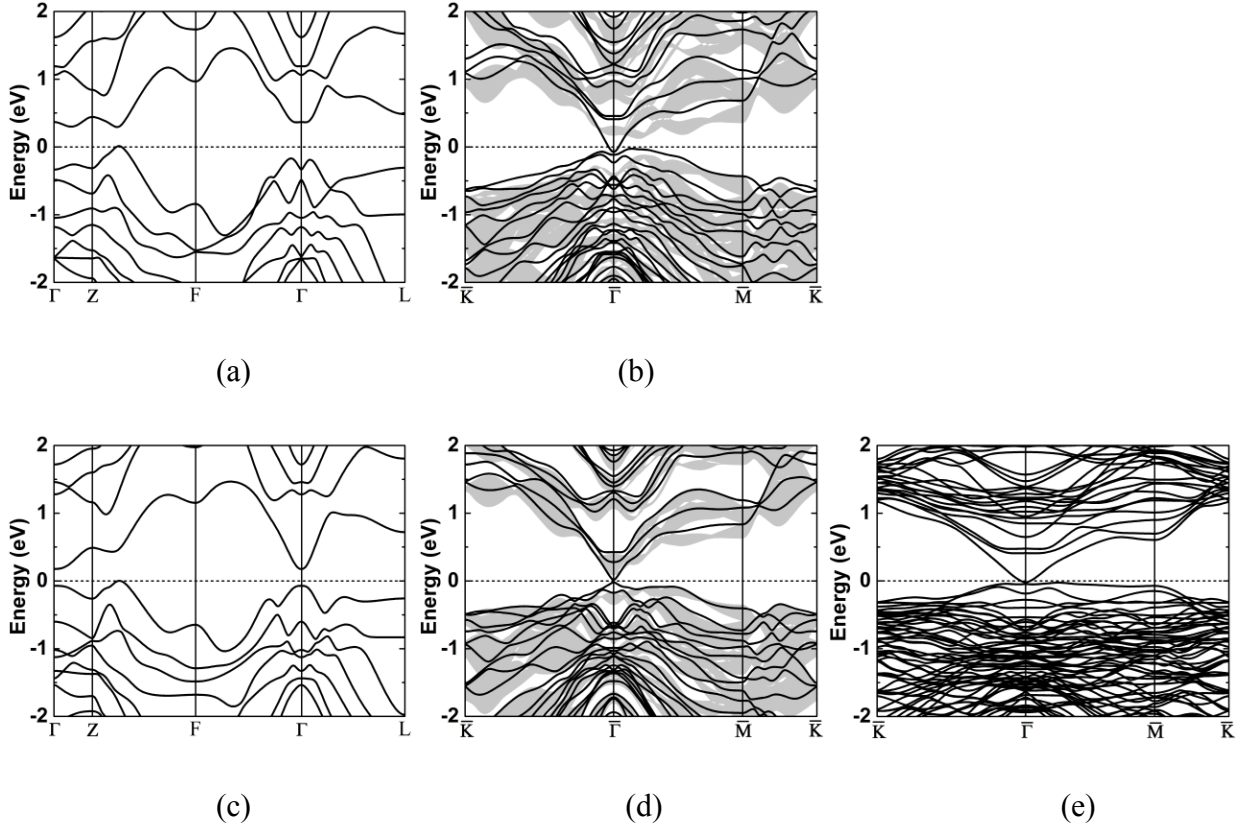

Fig. 2. Band structure with full SOC for $Bi_2Te_2Se$ and $Bi_2Se_2Te$. For $Bi_2Te_2Se$, we show (a) bulk dispersion and (b) slab dispersion along <111>. For $Bi_2Se_2Te$-I, (c) and (d) are similar to (a) and (b). For $Bi_2Se_2Te$-II, we show slab dispersion (e) along <111> (no bulk for clarity).

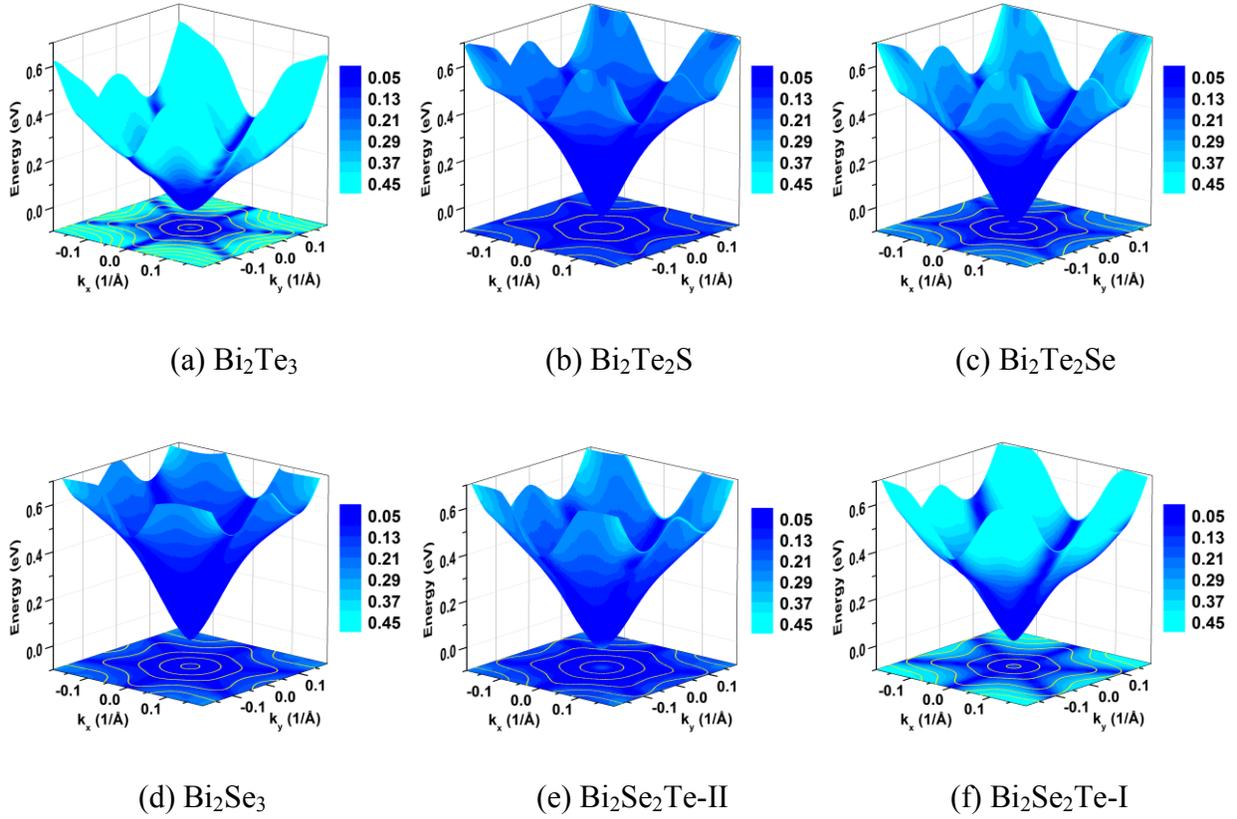

(a) Bi$_2$Te$_3$  (b) Bi$_2$Te$_2$S  (c) Bi$_2$Te$_2$Se

(d) Bi$_2$Se$_3$  (e) Bi$_2$Se$_2$Te-II  (f) Bi$_2$Se$_2$Te-I

Fig. 3. (Color online) Spin texture color-mapped on the conduction-band Dirac cone for the binary and ternary compounds studied (the hypothetical Bi$_2$Se$_2$Te-I shows the chemistry effect). Color indicates the amount of out-of-plane electronic spin moment in percent. Constant-energy contours are drawn on bottom.